\documentstyle[12pt]{article}
\newcounter{appendixc}
\newcounter{subappendixc}[appendixc]
\newcounter{subsubappendixc}[subappendixc]

\renewcommand{\appendix}[1] {\vspace*{0.6cm}
        \refstepcounter{appendixc}
        \setcounter{figure}{0}
        \setcounter{table}{0}
        \setcounter{equation}{0}
        \renewcommand{\thefigure}{\Alph{appendixc}.\arabic{figure}}
        \renewcommand{\thetable}{\Alph{appendixc}.\arabic{table}}
        \renewcommand{\theappendixc}{\Alph{appendixc}}
        \renewcommand{\theequation}{\Alph{appendixc}.\arabic{equation}}
        \noindent{\bf Appendix \theappendixc #1}\par\vspace*{0.4cm}}

\begin{document}        
\begin{titlepage}
\begin{flushright}
AMES-HET-97-3\\
March 1997
\end{flushright}
\vspace{0.1in}
\begin{center}
{\Large Dimension-six CP-violating operators of the third-family quarks\\
        and their effects at colliders}
\vspace{.2in}

  Jin Min Yang{\footnote{ On leave from Physics Department, 
Henan Normal University, China}},
  Bing-Lin Young

\vspace{.2in}
\it
     Department of Physics and Astronomy, and \\ 
     International Institute of Theoretical and Applied Physics,\\
     Iowa State University, Ames, Iowa 50011, USA
\rm
\end{center}
\vspace{4cm}

\begin{center} ABSTRACT\end{center}

We listed all possible dimension-six CP-violating 
$SU_c(3)\times SU_L(2)\times U_Y(1)$ invariant operators involving the 
third-family quarks, which can be generated by new physics at a higher 
energy scale. The expressions of these operators after electroweak 
symmetry breaking and the induced effective couplings $Wt\bar b$, $Xb\bar b$
and $Xt\bar t$ $(X=Z,\gamma,g,H)$ are also presented. 
We have evaluated sample contributions of these operators to 
CP-odd asymmetries of transverse polarization of top quark in
single top production at the upgraded Tevatron,
the similar effect in top-antitop pair production at the NLC,
and the CP-odd observables of momentum correlations among the top quark
decay products at the NLC.
The energy and luminosity sensitivity in probing these 
CP-violating new physics has also been studied. 
\vspace{.5in}

\end{titlepage}
\eject
\baselineskip=0.30in
\begin{center} {\Large 1. Introduction }\end{center}

It is widely believed that the Standard Model (SM) is only an effective theory
at electroweak scale and that some new physics should exist in higher energy 
regimes. Collider experiments have been searching for the new particles 
predicted by various models, but no direct 
signal has been observed. So, it is likely that
the new particles are too heavy to be detectable
at current colliders, and the only observable effects 
at energies not too far above the SM energy scale may be only 
in the form of new interactions. However, the new interactions will 
affect the couplings of third-family quarks, the 
Higgs and gauge bosons. In this spirit, the new physics effects can
be expressed as non-standard terms in an effective Lagrangian
involving the interactions of third-family quarks, the Higgs 
and gauge bosons. Before the electroweak symmetry breaking, we
can write the effective Lagrangian as
\begin{equation}\label{eq1}
{\cal L}_{eff}={\cal L}_0+\frac{1}{\Lambda^2}\sum_i C_i O_i
		         +{\cal O} (\frac{1}{\Lambda^4})
\end{equation}
where ${\cal L}_0$ is the SM Lagrangian, $\Lambda$ is the new physics
scale and $O_i$ are $SU_c(3)\times SU_L(2)\times U_Y(1)$ invariant
dimension-six operators,  and $C_i$ are
constants which represent the coupling strengths of $O_i$.
The expansion in Eq.(1) was first discussed in Ref. [1]. 
Recently, many authors further classified such CP-conserving operators 
and analysed their phenomenological implications at current and future
colliders[2-5]. 

As is well-known, for more than 30 years after the discovery of 
the CP-violating decays of the $K^0_L$ meson[6], the origin of this phenomenon 
remains a mystery. The SM gives a natural explanation for this phenomenon
assuming the existence of a phase in the Kobayashi-Maskawa mixing matrix[7].
In models beyond the SM, additional CP-violating effects can appear 
rather naturally and such non-standard CP-violations are necessary in order 
to account for baryogenesis[8]. In Ref.~[9], possible effects of non-SM
CP-violating interactions have been studied in detail 
in the form of momentum space representation and involving only weak bosons.
In this paper we will focus on CP-violation effects in the model-independent
effective Lagrangian approach. So we assume that the new physics terms in Eq.(\ref{eq1})
contain both CP-conserving and CP-violating  operators.

It has been shown[10] that the KM mechanism of CP-violation predicts
a negligibly small effect for the top quark in the SM, and  
thus the standard CP-violation effects in top production and decays will be
unobservable in collider experiments. Therefore, top quark system will be 
sensitive to new source of CP-violation
and may serve as a powerful probe to non-standard CP-violation
in association with new physics effects. Non-standard CP-violation 
in the top quark system as predicted by various new physics models and
the strategy for observing these effects have been studied by 
many authors[11-19]. Here we provide a model-independent 
study of all possible dimension-6 CP-violating operators which involve the 
third-family quarks and are invariant under the SM transformation.
The effects of these operators can be studied 
at future linear and hadron colliders, 
and thus their strengths can be constrained. We will 
evaluate some of the effects of these CP-violating operators 
at the Tevatron and the NLC. Any nonzero value of these CP
asymmetries will suggest the existence of new physics
as well as new CP-violation effects.

This paper is organized as follows. 
In Sec.2 we list all possible dimension-six CP-violating
$SU_c(3)\times SU_L(2)\times U_Y(1)$ invariant operators.
The expressions of these operators after electroweak gauge symmetry breaking
are given in Appendix A. In Sec.3 we give the induced CP-violating
effective couplings $Wt\bar b$, $Xb\bar b$ and $Xt\bar t$ $(X=Z,\gamma,g,H)$.
In Sec.4 we evaluate the contributions to some CP-odd quantities
 at the Tevatron and the NLC. And finally in Sec.5 we present the summary.
\vspace{.5cm}

\begin{center} {\Large 2. Dimension-six CP-violating gauge 
                          invariant operators}\end{center}
\vspace{.5cm}

We assume that the new physics in the quark sector resides in the third quark 
family. Although new physics can give rise to four-quark operators involving 
only the third family, such operators are not experimentally relevant here.
New physics may also occur in the gauge boson and Higgs sectors, they are 
not, however, our attention here.
Therefore, the operators we are interested in  are those containing 
third-family quarks coupling to gauge and Higgs bosons.

To restrict ourselves to the lowest order, we consider only tree diagrams
and to the order of $1/\Lambda^2$. Therefore, only one
vertex in a given diagram can contain anomalous couplings. Under these 
conditions, operators which are allowed to be related by the field equations
are not independent. As discussed in Ref.[5], to which we refer 
for the detail, the fermion and the Higgs boson equations of motion 
can be used but the equations of motion of the gauge bosons can not when 
writing down the operators in Eq.(1).

We assume all the operators $O_i$ to be Hermitian. Because of our assumption 
that the available energies are below the unitarity cuts of new-physics 
particles, no imaginary part can be generated by the new physics effect. 
Therefore the coefficients $C_i$ in Eq.(1) are real. 

Now we list all possible dimension-six CP-odd
$SU_c(3)\times SU_L(2)\times U_Y(1)$ invariant  operators
involving third-family quarks but no four-fermion interactions.
We follow the standard notation.

(1) Class 1 ( contain $t_R$ field )
\begin{eqnarray}
O_{t1}&=&i(\Phi^{\dagger}\Phi-\frac{v^2}{2})  
        \left [\bar q_L t_R\widetilde \Phi
        -\widetilde \Phi^{\dagger} \bar t_R q_L  \right ]\\
O_{t2}&=&  \left [\Phi^{\dagger}D_{\mu}\Phi
      +(D_{\mu}\Phi)^{\dagger}\Phi  \right ]\bar t_R \gamma^{\mu}t_R\\
O_{t3}&=&(\widetilde \Phi^{\dagger}D_{\mu}\Phi)(\bar t_R \gamma^{\mu}b_R)
         +(D_{\mu}\Phi)^{\dagger}\widetilde \Phi(\bar b_R \gamma^{\mu}t_R)\\
O_{Dt}&=&i  \left [(\bar q_L D_{\mu} t_R) D^{\mu}\widetilde \Phi
 -(D^{\mu}\widetilde\Phi)^{\dagger}(\overline {D_{\mu}t_R}q_L)  \right ]\\
\label{eq6}
O_{tW\Phi}&=&i  \left [(\bar q_L \sigma^{\mu\nu}\tau^I t_R) \widetilde\Phi
 -\widetilde\Phi^{\dagger}(\bar t_R \sigma^{\mu\nu}\tau^I q_L)  \right ]
 W^I_{\mu\nu}\\
O_{tB\Phi}&=&i\left [(\bar q_L \sigma^{\mu\nu} t_R) \widetilde\Phi
         -\widetilde\Phi^{\dagger}(\bar t_R \sigma^{\mu\nu} q_L)\right ]
          B_{\mu\nu}\\ \label{eq8}
O_{tG\Phi}&=&i\left [(\bar q_L \sigma^{\mu\nu}T^A t_R) \widetilde\Phi
         -\widetilde\Phi^{\dagger}(\bar t_R \sigma^{\mu\nu}T^A q_L)\right ]
          G^A_{\mu\nu}\\
O_{tG}&=&i\left [\bar t_R\gamma^{\mu}T^A D^{\nu}t_R
         -\overline{D^{\nu}t_R} \gamma^{\mu}T^A t_R\right ]
          G^A_{\mu\nu}\\
O_{tB}&=&i\left [\bar t_R \gamma^{\mu} D^{\nu}t_R
         -\overline{D^{\nu}t_R} \gamma^{\mu} t_R\right ]
          B_{\mu\nu}
\end{eqnarray}

(2) Class 2 (contain no $t_R$ field)
\begin{eqnarray}
O_{qG}&=&i\left [\bar q_L \gamma^{\mu}T^A D^{\nu}q_L
         -\overline{D^{\nu}q_L} \gamma^{\mu}T^A q_L\right ]
          G^A_{\mu\nu}\\
O_{qW}&=&i\left [\bar q_L \gamma^{\mu}\tau^I D^{\nu}q_L
         -\overline{D^{\nu}q_L} \gamma^{\mu}\tau^I q_L\right ]
          W^I_{\mu\nu}\\
O_{qB}&=&i\left [\bar q_L \gamma^{\mu} D^{\nu}q_L
         -\overline{D^{\nu}q_L} \gamma^{\mu} q_L\right ]
          B_{\mu\nu}\\
O_{bG}&=&i\left [\bar b_R \gamma^{\mu}T^A D^{\nu}b_R
         -\overline{D^{\nu}b_R} \gamma^{\mu}T^A b_R\right ]
          G^A_{\mu\nu}\\
O_{bB}&=&i\left [\bar b_R \gamma^{\mu} D^{\nu}b_R
         -\overline{D^{\nu}b_R} \gamma^{\mu}b_R\right ]
          B_{\mu\nu}\\
O_{\Phi q}^{(1)}&=&\left [\Phi^{\dagger}D_{\mu}\Phi
                   +(D_{\mu}\Phi)^{\dagger}\Phi\right ]
                   \bar q_L \gamma^{\mu}q_L\\
O_{\Phi q}^{(3)}&=&\left [\Phi^{\dagger}\tau^I D_{\mu}\Phi
                   +(D_{\mu}\Phi)^{\dagger}
                   \tau^I\Phi\right ]\bar q_L \gamma^{\mu}\tau^I q_L\\
O_{\Phi b}&=&\left [\Phi^{\dagger}D_{\mu}\Phi
             +(D_{\mu}\Phi)^{\dagger}\Phi\right ]
             \bar b_R \gamma^{\mu}b_R\\
O_{b1}&=&i(\Phi^{\dagger}\Phi-\frac{v^2}{2})\left [\bar q_L b_R\Phi
         -\Phi^{\dagger}\bar b_R q_L\right ]\\
O_{Db}&=&i\left [(\bar q_L D_{\mu} b_R) D^{\mu}\Phi
         -(D^{\mu}\Phi)^{\dagger}(\overline{D_{\mu}b_R}q_L)\right ]\\
\label{eq21}
O_{bW\Phi}&=&i\left [(\bar q_L \sigma^{\mu\nu}\tau^I b_R) \Phi
         -\Phi^{\dagger}(\bar b_R \sigma^{\mu\nu}\tau^I q_L)\right ]
          W^I_{\mu\nu}\\
O_{bB\Phi}&=&i\left [(\bar q_L \sigma^{\mu\nu} b_R) \Phi
         -\Phi^{\dagger}(\bar b_R \sigma^{\mu\nu} q_L)\right ]
          B_{\mu\nu}\\ \label{eq23}
O_{bG\Phi}&=&i\left [(\bar q_L \sigma^{\mu\nu}T^A b_R) \Phi
         -\Phi^{\dagger}(\bar b_R \sigma^{\mu\nu}T^A q_L)\right ]
          G^A_{\mu\nu}
\end{eqnarray}
Note that in $O_{t1}$ and $O_{b1}$ we  subtract the
vacuum expectation value, $v^2/2$, from $\Phi^{\dagger}\Phi$,
to avoid additional mass term for the third family quarks.

If we do not use the field equations of Higgs boson and the quarks,
we would have the following additional operators\\
(3) Class 3 
\begin{eqnarray}
O_{\overline{D}t}&=&i  \left [(\overline{D_{\mu}q_L} t_R) D^{\mu}\widetilde \Phi
         -(D^{\mu}\widetilde\Phi)^{\dagger}(\bar t_R D_{\mu}q_L)  \right ]\\
O_{\overline{D}b}&=&i\left [(\overline{D_{\mu}q_L} b_R) D^{\mu}\Phi
         -(D^{\mu}\Phi)^{\dagger}(\bar b_R D_{\mu}q_L)\right ]\\
O_{t\widetilde G}&=&\left [\bar t_R \gamma^{\mu}T^A D^{\nu}t_R
         +\overline{D^{\nu}t_R} \gamma^{\mu}T^A t_R\right ]
          \widetilde G^A_{\mu\nu}\\
O_{t\widetilde B}&=&\left [\bar t_R \gamma^{\mu} D^{\nu}t_R
         +\overline{D^{\nu}t_R} \gamma^{\mu} t_R\right ]
          \widetilde B_{\mu\nu}\\
O_{q\widetilde G}&=&\left [\bar q_L \gamma^{\mu}T^A D^{\nu}q_L
         +\overline{D^{\nu}q_L} \gamma^{\mu}T^A q_L\right ]
          \widetilde G^A_{\mu\nu}\\
O_{q\widetilde W}&=&\left [\bar q_L \gamma^{\mu}\tau^I D^{\nu}q_L
         +\overline{D^{\nu}q_L} \gamma^{\mu}\tau^I q_L\right ]
          \widetilde W^I_{\mu\nu}\\
O_{q\widetilde B}&=&\left [\bar q_L \gamma^{\mu} D^{\nu}q_L
         +\overline{D^{\nu}q_L} \gamma^{\mu} q_L\right ]
          \widetilde B_{\mu\nu}\\
O_{b\widetilde G}&=&\left [\bar b_R \gamma^{\mu}T^A D^{\nu}b_R
         +\overline{D^{\nu}b_R} \gamma^{\mu}T^A b_R\right ]
          \widetilde G^A_{\mu\nu}\\
O_{b\widetilde B}&=&\left [\bar b_R \gamma^{\mu} D^{\nu}b_R
         +\overline{D^{\nu}b_R} \gamma^{\mu}b_R\right ]
          \widetilde B_{\mu\nu},
\end{eqnarray}
where $\widetilde X_{\mu\nu}\equiv \frac{1}{2}\epsilon_{\mu\nu\lambda\rho} 
X^{\lambda\rho}$ for $X=G,B,W$ and $\epsilon_{\mu\nu\lambda\rho}$ the
anti-symmetric tensor. 
These Class 3 operators can be rewritten as
\begin{eqnarray}
O_{\overline{D}t}&=&-O_{Dt}-i[\bar q_L t_R D^2\widetilde \Phi
            -(D^2\widetilde \Phi)^{\dagger}\bar t_R q_L],\\
O_{\overline{D}b}&=&-O_{Db}-i[\bar q_L b_R D^2\Phi
                           -(D^2\widetilde \Phi)^{\dagger}\bar b_R q_L],\\
O_{x\widetilde B}&=&-O_{xB}-(\bar x_R \sigma_{\mu\nu}\not \! D x_R
+\overline{ \not \! D x_R}\sigma_{\mu\nu}x_R)B^{\mu\nu},~ (x=t,b),\\
O_{x\widetilde G}&=&-O_{xG}-(\bar x_R \sigma^{\mu\nu}T^a \not \! D x_R
+\overline{ \not \! D x_R}\sigma^{\mu\nu}T^a x_R)G^a_{\mu\nu},~(x=t,b),\\
O_{q\widetilde X}&=&O_{qX}
        +\bar q_L \sigma^{\mu\nu}X_{\mu\nu}\not \! D q_L
        +\overline{ \not \! D q_L}\sigma^{\mu\nu}X_{\mu\nu} q_L,
                 ~(X=B,G,W).
\end{eqnarray}
They become dependent upon the use of the field equations 
of the Higgs boson and the quarks,

It should also be noted that those CP-violating operators which are
obtained from Eqs.(\ref{eq6}-\ref{eq8}) and (\ref{eq21}-\ref{eq23}) by 
replacing the field tensors by their duals, $W^a_{\mu\nu}\rightarrow 
\widetilde W^a_{\mu\nu}$, etc., and changing the relative sign of the fermion
operators are not independent due to the identity
$\epsilon_{\mu\nu\lambda\rho}\sigma^{\lambda\rho}=2i\sigma^{\mu\nu}\gamma_5.$
For example, 
$ \left [(\bar q_L \sigma^{\mu\nu}\tau^I t_R) \widetilde\Phi
 +\widetilde\Phi^{\dagger}(\bar t_R \sigma^{\mu\nu}\tau^I q_L)  \right ]
\widetilde W^I_{\mu\nu}$ obtained from  Eq.(\ref{eq6}), is proportional to
 Eq.(\ref{eq6}).

The expressions of these CP-violating operators Eqs.(2-\ref{eq23})
after electroweak symmetry breaking are presented in Appendix A. 
Note that most of the operators clearly show the $U_{em}(1)$ gauge invariance.
But some of them do not manifest the electroweak gauge invariance 
straight forwardly, for example, $O_{Dt}$ in Eq.(\ref{Dt}).
We have checked that the operator gives indeed a $U_{em}(1)$
gauge invariant expression.

\vspace{1cm}

\begin{center} {\Large 3.~Effective Lagrangian for some couplings}
		\end{center}

We consider the contribution of CP-violating operators to top quark 
couplings $Wt\bar b$, $Zt\bar t$, $\gamma t\bar t$, $Ht\bar t$, $gt\bar t$  
and the bottom quark coupling $Zb\bar b, \gamma b\bar b$.
These couplings can be meaningfully investigated at LEP, Tevatron, NLC and LHC.
The status of the contributions of the dimension-six CP-violating operators 
to these couplings are showed in Table 1. 

Collecting all the relevant terms we get the CP-violating
effective couplings as
\begin{eqnarray}
\label{e38}
\tilde {\cal L}_{Wtb}&=&-i\frac{C_{\Phi q}^{(3)}}{\Lambda^2}\frac {g_2}{\sqrt 2}
      v^2 W^+_{\mu}(\bar t\gamma^{\mu}P_L b)
-i \frac{C_{t3}}{\Lambda^2} \frac{v^2}{2}\frac{g_2}{\sqrt 2} 
      W_{\mu}^+ (\bar t\gamma^{\mu}P_R b)\nonumber\\
& &-i\frac{C_{Dt}}{\Lambda^2}\frac{v}{\sqrt 2}\frac{g_2}{\sqrt 2}
     W_{\mu}^+ (i\partial^{\mu} \bar t)P_L b
     -i\frac{C_{Db}}{\Lambda^2}\frac{v}{\sqrt 2}\frac{g_2}{\sqrt 2} 
      W_{\mu}^+ \bar t P_R (i\partial^{\mu}b)\nonumber\\
& &-i\frac{C_{tW\Phi}}{\Lambda^2}\frac{v}{2}
        W^+_{\mu\nu}(\bar t\sigma^{\mu\nu}P_L b)
+i\frac{C_{bW\Phi}}{\Lambda^2}\frac{v}{2} 
      W^+_{\mu\nu}(\bar t \sigma^{\mu\nu}P_R b)\nonumber\\
& & +i\frac{C_{qW}}{\Lambda^2} \frac{1}{\sqrt 2}
        W^+_{\mu\nu}[\bar t\gamma^{\mu}P_L(\partial^{\nu}b)
                     -(\partial^{\nu}\bar t)\gamma^{\mu}P_L b]+h.c.\\
\tilde{\cal L}_{Zb\bar b}&=&
i(\frac{C_{bW\Phi}}{\Lambda^2}\frac{c_W}{\sqrt 2}
+\frac{C_{bB\Phi}}{\Lambda^2}\frac{v}{\sqrt 2}s_W) 
   Z_{\mu\nu}(\bar b \sigma^{\mu\nu}\gamma_5 b)\nonumber\\
& & +i(\frac{C_{qW}}{\Lambda^2}\frac{c_W}{2}
                  +\frac{C_{qB}}{\Lambda^2}s_W) 
   Z_{\mu\nu}(\bar b \gamma^{\mu}P_L \partial^{\nu}b
                      -\partial^{\nu}\bar b \gamma^{\mu}P_L b)\nonumber\\
& & +i\frac{C_{bB}}{\Lambda^2}s_W
    Z_{\mu\nu}(\bar b \gamma^{\mu}P_R \partial^{\nu}b
                -\partial^{\nu} \bar b\gamma^{\mu}P_R b)\nonumber\\
& &-i\frac{m_Z}{2}  
 Z^{\mu}\left [i(\bar b\gamma_5\partial_{\mu}b-\partial_{\mu}\bar b\gamma_5b)
   \frac{C_{Db}}{\Lambda^2}+i\partial_{\mu}(\bar b b)\frac{C_{Db}}{\Lambda^2}
      \right ]\\
\tilde{\cal L}_{\gamma b\bar b}&=&
 i (\frac{C_{qB}}{\Lambda^2}c_W-\frac{C_{qW}}{\Lambda^2}\frac{s_W}{2})
  A_{\mu\nu}(\bar b \gamma^{\mu}P_L \partial^{\nu}b
           -\partial^{\nu}\bar b \gamma^{\mu}P_L b)\nonumber\\
& & +i\frac{C_{bB}}{\Lambda^2}c_W
   A_{\mu\nu}(\bar b \gamma^{\mu}P_R \partial^{\nu}b
         -\partial^{\nu} \bar b\gamma^{\mu}P_R b)\nonumber\\
& &+i(\frac{C_{bB\Phi}}{\Lambda^2}c_W
     -\frac{C_{bW\Phi}}{\Lambda^2}\frac{s_W}{2})\frac{v}{\sqrt 2}
   A_{\mu\nu}(\bar b \sigma^{\mu\nu}\gamma_5 b)\\
\tilde{\cal L}_{Zt\bar t}&=&
   i\frac{C_{Dt}}{\Lambda^2}\frac{1}{\sqrt 2}\frac{m_Z}{2}
    Z^{\mu}[i\partial_{\mu}(\bar t t)]
+i\frac{C_{Dt}}{\Lambda^2}\frac{1}{\sqrt 2}\frac{m_Z}{2}
    Z^{\mu}(i\bar t\gamma_5\partial_{\mu}t-i\partial_{\mu}\bar t\gamma_5 t)
                  \nonumber\\
& &+i(\frac{C_{tB\Phi}}{\Lambda^2}s_W
 -\frac{C_{tW\Phi}}{\Lambda^2}\frac{c_W}{2})\frac{v}{\sqrt 2}
      Z_{\mu\nu}(\bar t\sigma^{\mu\nu}\gamma_5 t)
                  \nonumber\\
& &+i\frac{C_{tB}}{\Lambda^2}s_W
      Z_{\mu\nu}(\bar t\gamma^{\mu}P_R \partial^{\nu}t
         -\partial^{\nu} \bar t\gamma^{\mu}P_Rt)\nonumber\\
& &+i(\frac{C_{qB}}{\Lambda^2}s_W-\frac{C_{qW}}{\Lambda^2}\frac{c_W}{2}) 
      Z_{\mu\nu}(\bar t \gamma^{\mu}P_L \partial^{\nu}t
           -\partial^{\nu}\bar t \gamma^{\mu}P_L t)\\
\tilde{\cal L}_{\gamma t\bar t}&=&i(\frac{C_{tW\Phi}}{\Lambda^2}\frac{s_W}{2}
   +\frac{C_{tB\Phi}}{\Lambda^2}c_W)\frac{v}{\sqrt 2}
A_{\mu\nu}(\bar t\sigma^{\mu\nu}\gamma_5 t) \nonumber\\
& &  +i\frac{C_{tB}}{\Lambda^2}c_W
   A_{\mu\nu}(\bar t\gamma^{\mu}P_R \partial^{\nu}t
         -\partial^{\nu} \bar t\gamma^{\mu}P_Rt)\nonumber\\
& &
  +i(\frac{C_{qB}}{\Lambda^2}c_W+\frac{C_{qW}}{\Lambda^2}\frac{s_W}{2})
        A_{\mu\nu}(\bar t \gamma^{\mu}P_L \partial^{\nu}t
         -\partial^{\nu} \bar t \gamma^{\mu}P_L t)\\
\tilde{\cal L}_{H t\bar t}&=&i\frac{C_{t1}}{\Lambda^2} \frac{v^2}{\sqrt 2}H
                 (\bar t\gamma_5 t)
+i\frac{C_{Dt}}{\Lambda^2}
\frac{1}{2\sqrt 2}\partial^{\mu}H 
    \left [\partial_{\mu}(\bar t \gamma_5 t)
    +\bar t\partial_{\mu}t-(\partial_{\mu}\bar t) t
    \right ]\nonumber\\
& & -i\frac{C_{t2}}{\Lambda^2} v (i\partial^{\mu}H)
                 (\bar t\gamma_{\mu}P_R t)
-i(\frac{C^{(1)}_{\Phi q}}{\Lambda^2}-\frac{C^{(3)}_{\Phi q}}{\Lambda^2}) 
      v (i\partial^{\mu}H)(\bar t\gamma_{\mu}P_L t)\\
\tilde{\cal L}_{g t\bar t}&=&i\frac{C_{tG}}{\Lambda^2}
  \left [\bar t\gamma^{\mu}P_RT^A \partial^{\nu}t
         -\partial^{\nu} \bar t\gamma^{\mu}P_R T^A t \right ]G^A_{\mu\nu}
\nonumber\\
 & & +i\frac{C_{qG}}{\Lambda^2}
\left [\bar t\gamma^{\mu}P_L T^A \partial^{\nu}t
  -\partial^{\nu}\bar t \gamma^{\mu}P_L T^A t\right ] G^A_{\mu\nu} \nonumber\\
& & +i\frac{C_{tG\Phi}}{\Lambda^2}
  \frac{v}{\sqrt 2}(\bar t\sigma^{\mu\nu}\gamma_5T^A t)G^A_{\mu\nu}\\
\label{e45}
\tilde{\cal L}_{Hb\bar b}&=&
-i\frac{1}{\Lambda^2}(C_{\Phi q}^{(1)}+C_{\Phi q}^{(3)})
  v(i\partial_{\mu}H) \bar b\gamma^{\mu}P_Lb
-i\frac{C_{\Phi b}}{\Lambda^2}v(i\partial_{\mu}H)\bar b \gamma^{\mu}P_R b
\nonumber\\
& & +i\frac{C_{b1}}{\Lambda^2}\frac{v^2}{\sqrt 2}H(\bar b \gamma_5 b)
+i\frac{C_{Db}}{\Lambda^2}\frac{1}{2\sqrt 2}\partial^{\mu}H \left [
   \bar b\partial_{\mu}b-(\partial_{\mu}\bar b)b
   +\partial_{\mu}(\bar b\gamma_5 b) \right ],
\end{eqnarray}
where  $s_W\equiv \sin\theta_W$, $c_W\equiv \cos\theta_W$ and
$P_{L,R}\equiv(1\mp \gamma_5)/2$.
\vspace{1cm}

\begin{center} {\Large 4. The contributions to CP-odd quantities of top quark
			  at colliders} \end{center}
\vspace{.5cm}

Various experiments have been suggested to measure
CP-violating couplings of the top quark. They include CP-odd
quantities such as the polarization asymmetries[12-14] 
and CP-odd momentum correlations among the decay products[15,16].

In this section we will evaluate the contributions of some of the CP-violating 
new physics operators to these CP asymmetries. By taking 
individual operator as an example, we present numerical results to
show at what level of $C_i/\Lambda^2$ the CP-violating effect may be visible.
We will only consider the CP-odd operators listed
in Sec.3 and do not include their corresponding CP-even operators
whose phenomenologies are different and have been systematically analysed 
in Refs.[3-5].  Further more, we restrict ourselves to
the electroweak vertices,
i.e., $W t b$, $Z t \bar t$ and $\gamma t \bar t$. 
\vspace{.5cm}
\begin{center}

{\Large 4.1 Transverse polarization {\footnote{In this paper 
the transverse polarization direction is the one which is perpendicular
to the scattering plane.}} asymmetry of top quark
            in single top production at the Tevatron}\end{center}

The reaction $p\bar p\rightarrow t\bar bX$
at the Tevatron can be used to investigate several different
types of CP asymmetries[15]. 
The complicate coordinate representation of the effective Lagrangian
Eqs.(\ref{e38}-\ref{e45}) can be simplified in the momentum space when $t$ and $b$
are on-shell. 
The CP-violating contribution to the 
$W t b$ vertex Eqs.(\ref{e38}) can be written in the momentum space as
\begin{eqnarray}\label{e46}
\tilde {\cal L}_{Wtb}&=&i\frac{g_2}{\sqrt 2} W_{\mu}^+\bar t
 \left [F_L\gamma^{\mu}P_L+F_R\gamma^{\mu}P_R
-i\frac{G_L}{m_t}\sigma^{\mu\nu}k_{\nu}P_L
-i\frac{G_R}{m_t}\sigma^{\mu\nu}k_{\nu}P_R \right ]b
		\nonumber\\
& &-i\frac{g_2}{\sqrt 2} W_{\mu}^-\bar b
 \left [F_L\gamma^{\mu}P_L+F_R\gamma^{\mu}P_R
-i\frac{G_L}{m_t}\sigma^{\mu\nu}k_{\nu}P_R
-i\frac{G_R}{m_t}\sigma^{\mu\nu}k_{\nu}P_L\right ]t,
\end{eqnarray}
where $P_{L,R}\equiv(1\mp \gamma_5)/2$, $k=p_t+p_{\bar b}$,
and 
\begin{eqnarray}
F_L&=&\frac{v^2}{\Lambda^2}[-C_{\Phi q}^{(3)}
     +\frac{C_{Dt}}{2\sqrt 2}\frac{m_t}{v}],\\
G_L&=&\frac{v^2}{\Lambda^2}[\frac{C_{Dt}}{2\sqrt 2}\frac{m_t}{v}
+C_{tW\Phi}\frac{\sqrt 2}{g_2}\frac{m_t}{v}
-C_{qW}\frac{1}{g_2}\frac{m_t^2}{v^2}],\\
F_R&=&-\frac{v^2}{2\Lambda^2}[C_{t3}
     +\frac{C_{Db}}{\sqrt 2}\frac{m_t}{v}],\\
G_R&=&-\frac{v^2}{\Lambda^2}[\frac{C_{Db}}{2\sqrt 2}\frac{m_t}{v}
+C_{tW\Phi}\frac{\sqrt 2}{g_2}\frac{m_t}{v}],
\end{eqnarray}
We have neglected 
 the scalar and pseudoscalar couplings, $k_{\mu}$ and $k_{\mu}\gamma_5$, 
which, in the process $u\bar d\rightarrow W \rightarrow t\bar b$, 
give contributions proportional to the initial parton mass.
It should be pointed out that in contrast to Ref.[15], where the form factors 
$F_L$, etc., can be complex, form factors in Eq.(\ref{e46}) are all real because
$C^{(3)}_{\Phi q}$, etc., are real as noted in Sec.2 above.

The spin of the top quark allows three types
of CP-violating polarization asymmetries [15] in the single
top quark production via 
\begin{equation}
u+\bar d\rightarrow t+\bar b,~~
\bar u+d\rightarrow \bar t+ b.
\end{equation}
Introducing the coordinate system 
in the top quark ( or top antiquark) rest frame with the unit vectors
$\vec e_z \propto -\vec P_{\bar b}$, 
$\vec e_y \propto \vec P_u \times \vec P_{\bar b}$ and 
$\vec e_x=\vec e_y \times \vec e_z$, the transverse polarization asymmetry
is defined as 
\begin{equation}
A(\hat y)=\frac{1}{2}\left [\Pi(\hat y)-\bar \Pi(\hat y)\right ],
\end{equation}
where $\Pi(\hat y)$ and $\bar \Pi(\hat y)$ are, respectively, 
the polarizations of the top quark and top antiquark in the direction 
$\hat y$, arising from the interference of the SM and the CP-violating vertices.
Only the terms proportional to $P_L$ contribute.
The polarizations are given by
\begin{eqnarray}
\Pi(\hat y)&=&\frac{N_t(+\hat y)-N_t(-\hat y)}{N_t(+\hat y)+N_t(-\hat y)},
                                     \\
\bar \Pi(\hat y)&=&\frac{N_{\bar t}(+\hat y)-N_{\bar t}(-\hat y)}
                        {N_{\bar t}(+\hat y)+N_{\bar t}(-\hat y)},
\end{eqnarray}
where $N_t(\pm\hat y)$ $[N_{\bar t}(\pm\hat y)]$ is the number of
$t$($\bar t$) quarks polarized in the direction $\pm\hat y$.

The asymmetry $A(\hat y)$  is proportional to the
real part of the form factor $G_L$, which is given by[15]
\begin{equation}
A(\hat y)=\frac{3\pi}{4}\frac{(1-x)}{(2+x)\sqrt x}{\rm Re }~G_L,
\end{equation}
where $x=m_t^2/\hat s$.
This parton level asymmetry can be converted to the hadron level
asymmetry by folding in the structure functions. 
In the absence of an imaginary part $F_L$ makes no contribution
to polarization asymmetries.

Using the  CTEQ3L parton 
distribution functions[20] with $\mu=\sqrt {\hat s}$
and assuming $m_t=175$ GeV, we obtain the asymmetry as
\begin{equation}\label{Ay}
A(\hat y)=\left \{ 
\begin{array}{ll}
-0.41\frac{C_{qW}-2C_{tW\Phi}-g_2C_{Dt}/2}{(\Lambda/1 {\rm ~TeV})^2}
                                           &~~{\rm at}~\sqrt s=2~{\rm TeV}\\
-0.84\frac{C_{qW}-2C_{tW\Phi}-g_2C_{Dt}/2}{(\Lambda/1 {\rm ~TeV})^2}
                                           &~~{\rm at}~\sqrt s=4~{\rm TeV}
\end{array}\right.
\end{equation}

As analysed in Ref.[15], such an asymmetry of a few percent might be 
within the reach of experiment at the upgraded Tevatron with $\sqrt s=2$ TeV 
and an integrated luminosity 3-10 fb$^{-1}$.
As the results in Eq.(\ref{Ay}) show, the CP asymmetry caused by 
new physics will be more significant at higher energies, say
$\sqrt s=4$ TeV. Hence, if the collider can be further 
upgraded to 4 TeV and/or with increased luminosity[21], it can serve as a
more powerful tool for probing  CP-violating new physics.
It should be noted that the signal for this process is unobservable at the LHC
because of the large background from $t\bar t$ production and single
top production via $W$-gluon fusion[22]. 

Let's take $O_{qW}$ as an example.
If we assume an observable level of ten percent, we see from Eq.(\ref{Ay})
that the upgraded Tevatron will probe 
$\frac{C_{qW}}{(\Lambda/1 {\rm ~TeV})^2}$ to 1/4 and 1/8 for
$\sqrt s=2$ TeV and $\sqrt s=4$ TeV, respectively. 
This means that with a  new physics scale at the order of
1 TeV, the further upgraded Tevatron can probe the coupling strength 
down to the level of 0.1.
\vspace{.5cm}
\begin{center}

{\Large 4.2 Transverse polarization asymmetry of top quark 
            pair production at the NLC}\end{center}

From the polarizations of the top quark and top antiquark in 
 $e^+e^-\rightarrow t\bar t$, one can construct CP-odd quantities
which can be measured through the energy asymmetry of the charged leptons
in the $t$ and $\bar t$ decays as well as the up-down asymmetry of these leptons
with respect to the $t\bar t$ production plane[12,13].

Including both the SM couplings and new physics effects,  we can write
the $Vt\bar t$ $(V=Z,\gamma)$ vertices as
\begin{equation}\label{ver}
 \Gamma^{\mu}_{Vt\bar t}=i\frac{g}{2}\left [\gamma_{\mu}A_V
-\gamma_{\mu}\gamma_5 B_V+\frac{p_t^{\mu}-p_{\bar t}^{\mu}}{2}(C_V-iD_V\gamma_5)
\right ],
\end{equation}
where $p_t$ and $p_{\bar t}$ are the
momenta of the top quark and top antiquark. 
We neglect the scalar and pseudoscalar couplings, 
$k_{\mu}$ and $k_{\mu}\gamma_5$ with $k=p_t+p_{\bar t}$, since 
these terms give contributions proportional
to the electron mass. We note that some of these neglected terms are
needed to maintain the electromagnetic gauge invariance for the axial
vector couplings in Eq.(\ref{ver}).
The form factors can be written as
\begin{equation}
X_V=X_V^{\rm SM}+\delta X_V,~~(X=A,B,C,D ~{\rm and}~V=Z,\gamma), 
\end{equation}
where $X_V^{\rm SM}$ and $\delta X_V$ represent the SM and
the new physics contributions, respectively. 
In the SM, only $A_{\gamma,Z}$ and $B_Z$ exist at tree level. 
Beyond the tree level, all of them except the CP-violating form 
factor $D$ get contributions from loop diagrams.
The SM loop contribution to $D$ 
is completely negligible[10]. Since we are interested in CP-violation
effect, we neglect the SM loop contributions to 
all form factors. Thus we have 
\begin{eqnarray}
A^{\rm SM}_{\gamma}&=&\frac{4}{3}s_W,~
       A^{\rm SM}_Z=\frac{1}{2c_W}(1-\frac{8}{3}s_W^2),\\
B^{\rm SM}_{\gamma}&=&0,~B^{\rm SM}_Z=\frac{1}{2c_W},\\
C^{\rm SM}_{\gamma}&=&D^{\rm SM}_{\gamma}=C^{\rm SM}_Z=D^{\rm SM}_Z=0.
\end{eqnarray}
For new physics effects, only the form factor $D$ receives 
CP-violating contributions. Then we obtain
\begin{eqnarray}
\delta A_{\gamma,Z}&=&\delta B_{\gamma,Z}=\delta C_{\gamma,Z}=0,\\
\delta D_{\gamma}&=&-\frac{v}{\Lambda^2}\frac{4}{g}
           [(C_{qB}-C_{tB})\frac{c_Wm_t}{v}+C_{qW}\frac{s_Wm_t}{2v}
            -C_{tW\Phi}\frac{s_W}{\sqrt 2}-C_{tB\Phi}\sqrt 2 c_W],\\
\delta D_Z&=&\frac{v}{\Lambda^2}\frac{4}{g}
           [(C_{qB}-C_{tB})\frac{s_Wm_t}{v}-C_{qW}\frac{c_Wm_t}{2v}
            +C_{tW\Phi}\frac{c_W}{\sqrt 2}-C_{tB\Phi}\sqrt 2 s_W
            +C_{Dt}\frac{m_Z}{2\sqrt 2 v}].
\end{eqnarray}

The nonvanishing real parts of $D$ can give rise to the following asymmetry[14]
\begin{equation}
A_T=P_{\perp}\sin\alpha-\bar P_{\perp}\sin\bar\alpha,
\end{equation}
where $P_{\perp}\sin\alpha$ $(\bar P_{\perp}\sin\bar\alpha)$
is the degree of transverse polarization of the $t$ $(\bar t)$ quark
perpendicular to the scattering plane of
$e^+e^-\rightarrow t\bar t$. The scattering plane is defined to be
the $X$-$Z$ plane where the $+Z$ direction is the direction of electron
and the top-quark momentum has a positive $x$-component. 
The angle $\alpha$ depends on the top quark polarization direction
and its definition can be found in Appendix C 
of the first article of Ref. 14.
$P_{\perp}\sin\alpha$ and $\bar P_{\perp}\sin\bar\alpha$ are given by
\begin{equation}
P_{\perp}\sin\alpha =\frac{T_{\perp}}{G},~~
\bar P_{\perp}\sin\bar\alpha=\frac{\bar T_{\perp}}{G},
\end{equation}
where
\begin{eqnarray} 
G&=&~\vert (+-++)\vert^2+\vert (+-+-)\vert^2 
 +\vert (+--+)\vert^2+\vert (+---)\vert^2\nonumber\\
 & &+\vert (-+++)\vert^2+\vert (-++-)\vert^2 
 +\vert (-+-+)\vert^2+\vert (-+--)\vert^2,\\
T_{\perp}&=&2~{\rm Im}\left [
 (+-++)^*(+--+)+(+-+-)^*(+---)\right.\nonumber\\
 & &\left. ~~+(-+++)^*(-+-+)+(-++-)^*(-+--)\right ],\\
\bar T_{\perp}&=&2~{\rm Im}\left [
 (+-++)^*(+-+-)+(+--+)^*(+---)\right.\nonumber\\
 & &\left.~~+(-+++)^*(-++-)+(-+-+)^*(-+--)\right ].
\end{eqnarray}
Here the helicity amplitudes $(h_{e^-},h_{e^+},h_{t},h_{\bar t})$, where
$h_{e^-}=-,+$, etc., indicate respectively a left- and right-handed electron,
etc., are given by 
\begin{equation}
(h_{e^-},h_{e^+},h_{t},h_{\bar t})=2g^2E\left [ 
 \frac{(h_{e^-},h_{e^+},h_{t},h_{\bar t})_Z}{s-M_Z^2}
 + \frac{(h_{e^-},h_{e^+},h_{t},h_{\bar t})_{\gamma}}{s}\right ].
\end{equation}
The nonvanishing $(h_{e^-},h_{e^+},h_{t},h_{\bar t})_V$ $(V={\gamma},Z)$
can be found in Ref. 14 and are listed below:
\begin{eqnarray}
(-+--)_V&=&e_L^V\sin\theta_t(m_tA_V-K^2C_V+iEKD_V),\\
(-+-+)_V&=&-e_L^V(1+\cos\theta_t)(EA_V+KB_V),\\
(-++-)_V&=&e_L^V(1-\cos\theta_t)(EA_V-KB_V),\\
(-+++)_V&=&e_L^V\sin\theta_t(-m_tA_V+K^2C_V+iEKD_V),\\
(+---)_V&=&e_R^V\sin\theta_t(m_tA_V-K^2C_V+iEKD_V),\\
(+--+)_V&=&e_R^V(1-\cos\theta_t)(EA_V+KB_V),\\
(+-+-)_V&=&-e_R^V(1+\cos\theta_t)(EA_V-KB_V),\\
(+-++)_V&=&e_R^V\sin\theta_t(-m_tA_V+K^2C_V+iEKD_V),
\end{eqnarray}
where $\theta_t$ is the angle between the top quark and the electron,
 $E=\sqrt{s}/2$, $K=\sqrt{E^2-m_t^2}$ and 
$e^V_{L,R}$ are the form factors in $Ve^-e^+$ vertex 
$ig\gamma^{\mu}(e^V_LP_L+e^V_RP_R)$, which are given by
\begin{eqnarray}
e^Z_L&=&\frac{1}{c_W}(-\frac{1}{2}+s_W^2),~~
e^Z_R=\frac{1}{c_W}s_W^2,\\
e^{\gamma}_L&=&e^{\gamma}_R=-s_W.
\end{eqnarray}

As in the preceding subsection, we take the operator $O_{qW}$ as an example
to show the numerical results. Assuming the coupling strength $C_{qW}=0.1$, 
the asymmetry $A_T$ as a function of $\theta_t$ in the top pair
production at the NLC is plotted in Fig.1 and Fig.2 for $\sqrt s=500$ GeV
and $\sqrt s=1$ TeV, respectively.
Figure 1 shows that if the scale of new physics which generates the operator
$O_{qW}$ is below 1.5 TeV, the $A_T$ induced can exceed one 
percent. Comparing  Fig.1 with Fig.2, we find that the 
asymmetry $A_T$ for $\sqrt s=1$ TeV is larger than that for $\sqrt s=500$ GeV.
To see more clearly, we compare the values corresponding to
$\theta_t=120^{\circ}$
\begin{eqnarray*}
\begin{array}{l|cccc}
\Lambda  ({\rm TeV})~~~~& 0.5     & 1     & 1.5   & 2    \\ 
                        &         &       &       &      \\ \hline
A_T({\rm \%}) ~~~~      & -9.97~~ &-2.50~~&-1.11~~&-0.62\\
(\sqrt s=0.5~{\rm TeV})   ~~~~ &         &       &       &    \\ \hline
A_T({\rm \%}) ~~~~& -36.87~~&-9.34~~&-4.15~~&-2.34\\
(\sqrt s=1~  {\rm TeV})   ~~~~ &         &       &       &        
\end{array}
\end{eqnarray*}
Here we see that the $A_T$ for $\sqrt s=1$ TeV is four times
larger than that for $\sqrt s=500$ GeV.
But since the total event rate at a 1 TeV machine is about four times smaller
than a 500 GeV machine, the net effect is that a 1 TeV machine cannot
provide a better measurement unless it has a higher luminosity.
\vspace{.5cm}

\begin{center}
{\Large 4.3 Momentum correlations among the decay products of top quark
            at the NLC} \end{center}

In the process $e^+e^-\rightarrow \gamma^*, Z^*\rightarrow t\bar t$
with $t\rightarrow W^+b$ and $\bar t\rightarrow  W^- \bar b$,
some CP-odd momentum correlations among the decay products can be constructed
[15,16]. One of them, which is $CPT$-even and sensitive to the real
part of the dipole moment factor $D$ in Eq.(\ref{ver}), is 
\begin{equation}
O_1=(\vec p_b\times \vec p_{\bar b})\cdot \hat e_z,
\end{equation} 
where $\hat e_z$ is the unit vector along the incoming positron
beam direction. However, this observable is not sensitive to possible 
CP violation of the $t\bar b W$ vertex 
in the top quark decay [15,16]. Thus we consider only the CP-violating
new physics effects in the vertices $V t\bar t$($V=\gamma, Z$).
In terms of the expression Eq.(\ref{ver}), one gets the average value [17]
\begin{eqnarray}
\langle O_1 \rangle&=&-\frac{g}{48}sm_t(1-x)\epsilon^2\beta\Sigma^{-1}\left\{
\frac{1}{s^2}C^{\gamma\gamma}(v_e^{\gamma})^2v_t^{\gamma}{\rm Re}D_{\gamma}
			\right.\nonumber\\
& & +\frac{1}{s(s-m_Z^2)}C^{Z\gamma}v_e^{\gamma}v_e^Z 
(v_t^Z-\frac{\beta}{3}a_t^Z){\rm Re}D_{\gamma}\nonumber\\
& & +\frac{1}{s(s-m_Z^2)}C^{Z\gamma}v_e^{\gamma}v_e^Z 
           v_t^{\gamma}{\rm Re}D_Z \nonumber\\
& & \left.
+\frac{1}{(s-m_Z^2)^2}C^{ZZ}[(v_e^Z)^2+(a_e^Z)^2] 
(v_t^Z-\frac{\beta}{3}a_t^Z){\rm Re}D_Z\right\},
\end{eqnarray}
where 
\begin{eqnarray}
x&=&\frac{4m_t^2}{s},~\epsilon=1-\frac{m_W^2}{m_t^2},\nonumber\\
\beta&=&\frac{m_t^2-2m_W^2}{m_t^2+2m_W^2},\nonumber\\
C^{\gamma\gamma}&=&-p,~C^{Z\gamma}=\frac{a_e^Z}{v_e^Z}-p,\nonumber\\
C^{ZZ}&=&\frac{2a_e^Zv_e^Z}{(v_e^Z)^2+(a_e^Z)^2}-p,\nonumber\\
v_e^V&=&\frac{1}{2s_W}(e_L^V+e_R^V),~v_t^V=\frac{A_V}{2s_W},\nonumber\\
a_e^V&=&\frac{B_V}{2s_W},~a_t^V=\frac{1}{2s_W}(e_L^V-e_R^V),
\end{eqnarray}
and 
\begin{eqnarray}
\Sigma&=&
\frac{1}{s^2}(1+\frac{x}{2})(v_e^{\gamma})^2 (v_t^{\gamma})^2 
+\frac{2}{s(s-m_Z^2)}(1+\frac{x}{2})v_e^{\gamma}v_t^{\gamma}
(v_e^Z-pa_e^Z)v_t^Z\nonumber\\
& & +\frac{1}{(s-m_Z^2)^2}\left [(v_e^Z)^2+(a_e^Z)^2-2p v_e^Z a_e^Z\right ] 
\left [(1+\frac{x}{2})(v_t^Z)^2+(1-x)(a_t^Z)^2\right ].
\end{eqnarray}
In the above equations, $s$ is the center-of-mass energy squared and 
$p$ is the degree of longitudinal polarization of the initial
electron with $p=\pm 1$ corresponding to the right- and left-handed helicities, 
respectively.
Note that in our analyses we neglect both the radiative corrections to 
the couplings $V e^+e^-$ ($V=\gamma, Z$) and the 
electron mass, thus only
the left-right and right-left combinations
{\footnote{ Hard collinear emission of a photon from the electron and
positron beams can flip helicities. This gives rise to non-zero CP-odd
correlations even in the absence of CP-violating interactions and this
background should be subtracted. However, as analysed in Ref.[17],
there will be no such background at tree level for $\langle O_1 \rangle$.}}
 of electron and positron
helicities couple to the $\gamma$ and $Z$.  

Again we take the operator $O_{qW}$ as an example to show some 
results. The values of $\langle O_1 \rangle$ for
different polarizations of the electron beam with 
new physics scale of 1 TeV and coupling strength of unity
are found to be
\begin{eqnarray*}
\begin{array}{l|cccc}
                        & e^+e^-_L&e^+e^-_R&e^+e^- \\ 
                        &         &        &       \\ \hline
\langle O_1 \rangle~[({\rm GeV})^2]      & -36.7~~ &-1.0~~  &-25.5  \\
(\sqrt s=0.5~{\rm TeV}) &         &        &       \\ \hline
\langle O_1 \rangle~[({\rm GeV})^2]     &-272.3~~ &-1.7~~  &-183.4 \\
(\sqrt s=1~  {\rm TeV}) &         &        &              
\end{array}
\end{eqnarray*}
Here we find that the left-polarized electron beam yields the most
significant results for $\langle O_1 \rangle$ and in this case the result in
a 1 TeV accelerator is eight times larger than a 500 GeV accelerator.
In the following analyses we will only consider the 
left-polarized electron beam.
 
Now we compare the value of $\langle O_1 \rangle$ with the expected variance
$\langle O_1^2\rangle$ to see what luminosity is needed for the observation
to be statistically  significant.
To observe a deviation from the SM expectation with better than
one standard deviation ( at the 68\% confidence level), we need
\begin{equation}
\vert \langle O_1 \rangle\vert \geq \sqrt { \frac {\langle O_1^2\rangle}{{\cal L}\sigma\kappa}},
\end{equation}
where ${\cal L}$ is the integrated luminosity, $\kappa$ is 
the overall $b$- and $W$-tagging efficiency. The variance
$\langle O_1^2\rangle$ and the production cross section $\sigma$ at lowest order 
are given by[17]
\begin{eqnarray}
\sigma&=&4\pi\alpha^2 s \sqrt{1-x} \Sigma,\\
\langle O_1^2\rangle&=&\frac{sm_t^2\epsilon^4}{2880}\Sigma^{-1}\left\{
\frac{1}{s^2}(v_e^{\gamma})^2 (v_t^{\gamma})^2[24+2x-11x^2+4\beta^2(1-x)^2]
		\right.\nonumber\\
& & +\frac{2}{s(s-m_Z^2)}v_e^{\gamma}v_e^Z \left [v_t^{\gamma}v_t^Z
          \left (24+2x-11x^2+4\beta^2(1-x)^2\right )
    -2v_t^{\gamma}a_t^Z(1-x)(6-x)\beta\right ]\nonumber\\
& & +\frac{1}{(s-m_Z^2)^2}[(v_e^Z)^2+(a_e^Z)^2] 
\left [ (v_t^Z)^2\left (24+2x-11x^2+4\beta^2(1-x)^2\right )\right.\nonumber\\
& & \left.\left.+(a_t^Z)^2\left (24-14x-4\beta^2(1-x)\right )(1-x)
    -4v_t^Za_t^Z(1-x)(6-x)\beta\right ]\right\}.
\end{eqnarray}
For a negative helicity electron beam considered in our analyses, 
the production rate is 
\begin{equation}\label{rate}
\sigma(e^+e^-_L\rightarrow t\bar t)=\left \{
\begin{array}{ll}
775~ {\rm fb} & {~~\rm for~}\sqrt s=500~ {\rm GeV},\\
232~ {\rm fb} & {~~\rm for~}\sqrt s=1 ~  {\rm TeV}.
\end{array}\right.
\end{equation}

Assuming the coupling strength of the order of unity and an
overall $b$- and $W$-tagging 
efficiency of 50\%, then the 
luminosity required to observe the CP-violating effects of $O_{qW}$
at 68\% confidence level is found to be
\begin{equation}
{\cal L}=\left \{ 
\begin{array}{ll}
25\frac{(\Lambda/1 {\rm ~TeV})^4}{C_{qW}^2}~{\rm fb}^{-1} 
                                            &~~{\rm at}~\sqrt s=0.5~{\rm TeV}\\
8\frac{(\Lambda/1 {\rm ~TeV})^4}{C_{qW}^2}~{\rm fb}^{-1} 
                                              &~~{\rm at}~\sqrt s=1~{\rm TeV}
\end{array}\right.
\end{equation}
So, if the new physics scale is 1 TeV, we need a luminosity of 
$100$ fb$^{-1}$ ($30$ fb$^{-1}$) to probe the coupling strength 
$C_{qW}$ down to 0.5 with a confidence level of 68\% at 
$\sqrt s=500$ GeV (1 TeV).
If a conservative overall $b$- and $W$-tagging efficiency of 10\%
is assumed, the required luminosity will be increased by a factor of 5.
If a confidence level of 99.7\%  is assumed, 
the required luminosity will be increased by a factor of 9.

From the above results we find that for the same luminosity
a 1 TeV collider can do a better measurement than a 500 GeV collider.
This is due to the fact that the size of $\langle O_1 \rangle$ at 
$\sqrt s=1$ TeV is eight times larger than at $\sqrt s=500$ GeV,
while the production rate at $\sqrt s=1$ TeV is only about four 
times smaller than at $\sqrt s=500$ GeV. Thus the net effect is that 
a 1 TeV accelerator can do a better measurement than a 500 GeV accelerator.

\begin{center} {\Large 5. Summary} \end{center}

In this paper we listed all possible dimension-six CP-violating 
$SU_c(3)\times SU_L(2)\times U_Y(1)$ invariant operators involving the 
third-family quarks, which may be generated by new physics at a higher scale.
The expressions of these operators after the electroweak symmetry breaking
and the induced effective couplings for $Wt\bar b$, $Vb\bar b$
and $Vt\bar t$ $(V=Z,\gamma,g,H)$ were presented. 

The contributions of some of these operators to the CP-odd asymmetries of
the transverse polarization of top quark and top antiquark in
single top production at the Tevatron and top pair production
at the NLC are evaluated.
The numerical results showed that if the new physics scale
is around 1 TeV, then both colliders can be used to probe the coupling strength
to 0.1 provided that the asymmetry of the transverse polarization 
can be measured at a level of a few percent.

We also calculated the effects on a CP-odd observable, which involves momentum 
correlations among the decay products of the top quark,  at the NLC
and studied the dependence on the energy and luminosity of the NLC.
 We found that with a luminosity
of 100 fb$^{-1}$, a 500 GeV accelerator can probe the coupling strength
to 0.5, assuming that the new physics scale is of the order of 1 TeV. 
Achieving the same measurement, we need a luminosity of 30 fb$^{-1}$ at a 
1 TeV accelerator.
\vspace{1cm}

\begin{center}{\Large  Acknowledgement}\end{center}

J.M.Y. thanks C.-P.Yuan for discussions.
This work was supported in part by the U.S. Department of Energy, Division
of High Energy Physics, under Grant No. DE-FG02-94ER40817.
\vspace{.5cm}

\appendix{~~~CP-violating operators after electroweak symmetry breaking}

(1) Class 1
\begin{eqnarray}
O_{t1}&=&\frac{1}{2\sqrt 2}H(H+2v)(H+v)(\bar ti\gamma_5 t)\\
O_{t2}&=&(H+v)\partial^{\mu}H(\bar t_R \gamma_{\mu} t_R)\\
O_{t3}&=&i\frac{1}{2\sqrt 2}g_2 (H+v)^2\left [
          -W_{\mu}^+(\bar t_R\gamma^{\mu}b_R)
         +W_{\mu}^- (\bar b_R\gamma^{\mu} t_R)\right ]\\ \label{Dt}
O_{Dt}&=&i\frac{1}{2\sqrt 2}\partial^{\mu}H \left [\bar t\partial_{\mu}t
                   -(\partial_{\mu}\bar t)t+\partial_{\mu}(\bar t\gamma_5 t)
		   -i\frac{4}{3}g_1 B_{\mu}\bar t t\right ]\nonumber\\
& & -\frac{1}{4\sqrt 2}g_Z (H+v)Z^{\mu}\left [\partial_{\mu}
(\bar t t)+\bar t\gamma_5\partial_{\mu}t-(\partial_{\mu}\bar t)\gamma_5 t
-i\frac{4}{3}g_1 B_{\mu}\bar t \gamma_5 t\right ]\nonumber\\
& & +\frac{1}{2}g_2 (H+v)W_{\mu}^- \left [\bar b_L\partial^{\mu} t_R
                 -i\frac{2}{3}g_1 B^{\mu}\bar b_L  t_R\right ]\nonumber\\
& & +\frac{1}{2}g_2 (H+v)W_{\mu}^+ \left [(\partial^{\mu} \bar t_R)b_L
                 +i\frac{2}{3}g_1 B^{\mu}\bar t_R  b_L\right ]\\
O_{tW\Phi}&=&i\frac{1}{2\sqrt 2}(H+v)(\bar t\sigma^{\mu\nu}\gamma_5 t)
      \left [W^3_{\mu\nu}-ig_2(W^+_{\mu}W^-_{\nu}
      -W^-_{\mu}W^+_{\nu})\right ]\nonumber\\
& & +i\frac{1}{2}(H+v)(\bar b_L\sigma^{\mu\nu} t_R)
      \left [W^-_{\mu\nu}-ig_2(W^-_{\mu}W^3_{\nu}
      -W^3_{\mu}W^-_{\nu})\right ]\nonumber\\
& & -i\frac{1}{2}(H+v)(\bar t_R\sigma^{\mu\nu} b_L)
      \left [W^+_{\mu\nu}-ig_2(W^3_{\mu}W^+_{\nu}-W^+_{\mu}W^3_{\nu})\right ]\\
O_{tB\Phi}&=&i\frac{1}{\sqrt 2}(H+v)
            (\bar t\sigma^{\mu\nu}\gamma_5 t)B_{\mu\nu}\\
O_{tG\Phi}&=&i\frac{1}{\sqrt 2}(H+v)(\bar t\sigma^{\mu\nu}\gamma_5 T^A t)
				G^A_{\mu\nu}\\
O_{tG}&=&i\left [\bar t_R\gamma^{\mu}T^A \partial^{\nu}t_R
         -\partial^{\nu} \bar t_R\gamma^{\mu}T^A t_R\right ]G^A_{\mu\nu}
\nonumber\\
& & +g_s \bar t_R\gamma^{\mu}\left \{G^{\nu},G_{\mu\nu}\right \}t_R
    +\frac{4g_1}{3}\bar t_R\gamma^{\mu}G_{\mu\nu}B^{\nu}t_R\\
O_{tB}&=&i\left [\bar t_R\gamma^{\mu} \partial^{\nu}t_R
    -\partial^{\nu} \bar t_R\gamma^{\mu}t_R\right ]B_{\mu\nu}
    +2g_s \bar t_R\gamma^{\mu}G^{\nu}t_R B_{\mu\nu}
    +\frac{4}{3}g_1\bar t_R\gamma^{\mu}t_R B_{\mu\nu}B^{\nu}
\end{eqnarray}
(2) Class 2
\begin{eqnarray}
O_{qG}&=&i\left [\bar q_L \gamma^{\mu}T^A \partial^{\nu}q_L
         -\partial^{\nu}\bar q_L \gamma^{\mu}T^A q_L\right ]
          G^A_{\mu\nu}\nonumber\\
& & +g_s \bar q_L\gamma^{\mu}\left \{G^{\nu},G_{\mu\nu}\right \}q_L
    +2g_2\bar q_L\gamma^{\mu}W^{\nu}G_{\mu\nu}q_L
    +\frac{1}{3}g_1 \bar q_L\gamma^{\mu}G_{\mu\nu}B^{\nu}q_L\\
O_{qW}&=&\frac{i}{2}W^3_{\mu\nu}\left [\bar t_L\gamma^{\mu}\partial^{\nu}t_L
-\partial^{\nu}\bar t_L\gamma^{\mu}t_L
-\bar b_L\gamma^{\mu}\partial^{\nu}b_L
+\partial^{\nu}\bar b_L\gamma^{\mu}b_L\right ]	\nonumber\\
& & +\frac{i}{\sqrt 2}\left [W^+_{\mu\nu}(\bar t_L\gamma^{\mu}\partial^{\nu}b_L
-\partial^{\nu}\bar t_L\gamma^{\mu}b_L)
+W^-_{\mu\nu}(\bar b_L\gamma^{\mu}\partial^{\nu}t_L
-\partial^{\nu}\bar b_L\gamma^{\mu}t_L)	\right ]\nonumber\\
& & +g_2\bar q_L \gamma^{\mu} \left [W_{\mu},W_{\nu}\right ]\partial^{\nu}q_L
  -g_2\partial^{\nu}\bar q_L \gamma^{\mu} 
  \left [W_{\mu},W_{\nu}\right ]q_L\nonumber\\
& &+2g_s \bar q_L \gamma^{\mu} G^{\nu}W_{\mu\nu}q_L
   +\frac{1}{2}g_2 (\vec W_{\mu\nu}\cdot \vec W^{\nu})
   \bar q_L \gamma^{\mu} q_L
  +\frac{1}{3}g_1 B^{\nu} \bar q_L \gamma^{\mu}W_{\mu\nu} q_L\\
O_{qB}&=&iB_{\mu\nu}\left [\bar q_L\gamma^{\mu} \partial^{\nu}q_L
   -\partial^{\nu} \bar q_L\gamma^{\mu}q_L
   -2i\bar q_L\gamma^{\mu}(g_sG^{\nu}+g_2 W^{\nu}
   +\frac{1}{6}g_1B^{\nu})q_L\right ]\\
O_{bG}&=&i\left [\bar b_R \gamma^{\mu}T^A \partial^{\nu}b_R
         -\partial^{\nu}\bar b_R \gamma^{\mu}T^A b_R\right ]G^A_{\mu\nu}
\nonumber\\ 
& &+g_s \bar b_R\gamma^{\mu}\left \{G^{\nu},G_{\mu\nu}\right \}b_R
    -\frac{2g_1}{3}\bar b_R\gamma^{\mu}G_{\mu\nu}B^{\nu}b_R\\
O_{bB}&=&i\left [\bar b_R\gamma^{\mu} \partial^{\nu}b_R
         -\partial^{\nu} \bar b_R\gamma^{\mu}b_R\right ]B_{\mu\nu}
  +2g_s \bar b_R\gamma^{\mu}G^{\nu}b_R B_{\mu\nu}
    -\frac{2}{3}g_1\bar b_R\gamma^{\mu}b_R B_{\mu\nu}B^{\nu}\\
O_{\Phi q}^{(1)}&=&(H+v)\partial_{\mu}H \left [\bar t_L\gamma^{\mu}t_L
          +\bar b_L\gamma^{\mu}b_L\right ] \\
O_{\Phi q}^{(3)}&=&-O_{\Phi q}^{(1)}+2(H+v)\partial_{\mu}H 
        \bar b_L\gamma^{\mu}b_L\nonumber\\
& & -\frac{i}{\sqrt 2}g_2(H+v)^2(W^+_{\mu}\bar t_L\gamma^{\mu}b_L
		-W^-_{\mu}\bar b_L\gamma^{\mu}t_L)\\
O_{\Phi b}&=&(H+v)\partial_{\mu}H \bar b_R \gamma^{\mu}b_R\\
O_{b1}&=&\frac{1}{2\sqrt 2}H(H+v)(H+2v)\bar b i\gamma_5 b\\
O_{Db}&=&i\frac{1}{2\sqrt 2}\partial^{\mu}H \left [
   \bar b\partial_{\mu}b-(\partial_{\mu}\bar b)b
   +\partial_{\mu}(\bar b\gamma_5 b)
   +i\frac{2}{3}g_1B_{\mu}(\bar b b)\right ]\nonumber\\
& & +\frac{1}{4}g_Z (H+v)Z^{\mu}\left [\partial_{\mu}(\bar b b)
    +\bar b\gamma_5\partial_{\mu}b-(\partial_{\mu}\bar b)\gamma_5 b
    +\frac{2}{3}g_1B_{\mu}(\bar bi\gamma_5 b)\right ]\nonumber\\
& & +\frac{g_2}{2}(H+v)\left [W_{\mu}^+ (\bar t_L \partial_{\mu}b_R
    +\frac{i}{3}g_1 B_{\mu} \bar t_L  b_R)
+W_{\mu}^- ( \partial_{\mu}\bar b_R t_L-
                 \frac{i}{3}g_1 B_{\mu} \bar b_R t_L)\right ]\\
O_{bW\Phi}&=&i\frac{1}{2}(H+v)\left [W^+_{\mu\nu}(\bar t_L \sigma^{\mu\nu} b_R)
    -W^-_{\mu\nu}(\bar b_R \sigma^{\mu\nu} t_L)
    -\frac{1}{\sqrt 2}W^3_{\mu\nu}
    (\bar b \sigma^{\mu\nu}\gamma_5 b)\right.\nonumber\\
& & +ig_2(W^+_{\mu}W^3_{\nu}-W^3_{\mu}W^+_{\nu})(\bar t_L \sigma^{\mu\nu} b_R)
    +ig_2(W^-_{\mu}W^3_{\nu}-W^3_{\mu}W^-_{\nu})(\bar b_R \sigma^{\mu\nu} t_L)
\nonumber\\
& &\left. +i\frac{g_2}{\sqrt 2} (W^+_{\mu}W^-_{\nu}-W^-_{\mu}W^+_{\nu})
      (\bar b \sigma^{\mu\nu}\gamma_5 b)\right ]\\    
O_{bB\Phi}&=&\frac{i}{\sqrt 2}(H+v) B_{\mu\nu}
            (\bar b \sigma^{\mu\nu}\gamma_5 b)\\
O_{bG\Phi}&=&\frac{i}{\sqrt 2}(H+v) G^A_{\mu\nu}
            (\bar b \sigma^{\mu\nu}\gamma_5 T^A b)
\end{eqnarray}
\vspace{1cm}

{\LARGE References}
\vspace{0.3in}
\begin{itemize}
\begin{description}
\item[{\rm [1]}] C. J. C. Burgess and H. J. Schnitzer, 
                 Nucl. Phys. B228, 454 (1983);\\
                 C. N. Leung, S. T. Love and S. Rao, Z. Phys. C31, 433 (1986);\\
                 W. Buchmuller and D. Wyler, Nucl. Phys. B268, 621 (1986). 
\item[{\rm [2]}] K. Hagiwara, S. Ishihara, R. Szalapski and D. Zeppenfeld,
                 Phys. Lett. B283, 353 (1992); 
                                   Phys. Rev. D48, 2182 (1993). 
\item[{\rm [3]}] G. J. Gounaris, F. M. Renard and C. Verzegnassi, 
                 Phys. Rev. D52, 451 (1995);\\ 
          R. D. Peccei and X. Zhang, Nucl. Phys. B337, 269  (1990);\\
          R. D. Peccei, S. Peris and X. Zhang, Nucl. Phys. B349, 305  (1991);\\
          C. T. Hill and S. Parke, Phys. Rev. D49, 4454  (1994);\\
          D. Atwood, A. Kagan and T. Rizzo, Phys. Rev. D52, 6264  (1995); \\
       D. O. Carlson, E. Malkawi and C.-P. Yuan, Phys. Lett. B337,145 (1994);\\
       H. Georgi, L. Kaplan, D. Morin and A. Shenk, Phys. Rev. D51, 3888
              							      (1995);\\ 
       T. Han, R. D. Peccei and X. Zhang, Nucl. Phys. B454, 527 (1995);\\
       X. Zhang and B.-L. Young,  Phys. Rev. D51, 6564 (1995);\\
       E. Malkawi and T. Tait, Michigan State Preprint, MSUHEP-51116 (1995);\\
       S. Dawson and G. Valencia, Phys. Rev. D53, 1721  (1996);\\
       T. G. Rizzo,  Phys. Rev. D53, 6218 (1996);\\
       P. Haberl, O. Nachtman and A. Wilch,  Phys. Rev. D53, 4875 (1996);\\
      T. Han, K. Whisnant and B.-L. Young and X. Zhang, 
                                  Phys.Lett.B385, 311(1996);\\
      T. Han, K. Whisnant and B.-L. Young and X. Zhang, hep-ph/9603247, 
                                  to be published in Phys.Rev.D;\\
      K. Hagiwara, T. Hatsukano, S. Ishihara and R. Szalapski, hep-ph/9612268;\\
      G. J. Gounaris, J. Layssac and F. M. Renard, hep-ph/9612335, 
                                  to be published in Phys.Rev.D;\\
      A. Data and X. Zhang, Phys. Rev. D55, 2530 (1997);\\
      B.-L. Young, 
         {\it Proceedings of the International Symposium on Heavy Flavor
                    and Electroweak Theory}, August 1995, ed by C.-H.Chang and
                    C.-S. Huang (World Scientific, 1996).  
\item[{\rm [4]}] G. J. Gounaris, D. T. Papadamou and F. M. Renard,
                 hep-ph/9609437. 
\item[{\rm [5]}] K. Whisnant, J. M. Yang, B.-L. Young and X. Zhang, 
                 hep-ph/9702305.
\item[{\rm [9]}] K. Hagiwara, R. D. Peccei and D. Zeppanfeld, 
                 Nucl. Phys. B282, 253 (1987);\\
	         D. London, Phys. Rev. D45, 3186 (1991);\\
       X. G. He, J. P. Ma, B. H. J. Mckellar, Phys. Lett. B304, 285 (1993);\\
       D. Choudhury and S. S. Rindani, Phys. Lett. B335, 198 (1994);\\
       G. Belanger and G. Couture, Phys. Rev. D49, 5720 (1994);\\
       D. Chang, W. Y. Keung and P. Pal, Phys. Rev. D51, 1326 (1995). 
\item[{\rm [6]}] J. Christenson, J. W. Cronin, V. L. Fitch and R. Turlay, Phys. 
                 Rev. Lett. 13, 138  (1964). 
\item[{\rm [7]}] K. Kobayashi, T. Maskawa, Prog. Theor. Phys. 49, 652  (1973). 
\item[{\rm [8]}] For a review see A. G. Cohen, D. B. Kaplan and A. E. Nelson,
                  Annu. Rev. Nucl. Phys. 43, 27  (1993). 
\item[{\rm [10]}] C. Jarlskog, Phys. Rev. D35, 1685 (1987);\\
                 C. Jarlskog and R. Stora, Phys. Lett. B208, 268 (1988);\\
       G. Eilam, J. Hewett and A. Soni, Phys. Rev. Lett. 67, 1979 (1991);\\
       B. Grzadkowski, Phys. Lett. B319, 526 (1993). 
\item[{\rm [11]}] For reviews see, for example, \\
                 C.-P. Yuan, Mod. Phys. Lett. A10,627 (1995);\\
                 D. Atwood and A. Soni, hep-ph/9609418, to appear in
                 {\it Proceedings of the 28th Int. Conf. on HEP}, Warsaw 
                 (July 1996).
\item[{\rm [12]}] G. L. Kane, J. Pumplin and W. Repko, Phys. Rev. Lett. 41,
                  1689 (1978);\\
       A. Devoto, G. L. Kane, J. Pumplin and W. Repko, Phys. Rev. Lett. 43,
       1062  (1979); Phys. Rev. Lett. 43, 1540  (1979); 
       Phys. Lett. B90, 436 (1980). 
\item[{\rm [13]}] J. F. Donoghue anf G. Valencia, 
                  Phys. Rev. Lett. 58, 451  (1987);\\
	 C. R. Schmidt and M. E. Peskin, Phys. Rev. Lett. 69, 410  (1992);\\
         D. Chang, W.-Y. Keung and I. Phillips, Nucl. Phys. B408, 286 (1993). 
\item[{\rm [14]}] G. L. Kane, G. A. Ladinsky and C.-P. Yuan,
                  Phys. Rev. D45, 124 (1992);\\
	          C.-P. Yuan,Phys. Rev. D45, 782  (1992);\\
                  G. A. Ladinsky and C.-P. Yuan,Phys. Rev. D49, 4415 (1994). 
\item[{\rm [15]}] D. Atwood, S. B. Shalom, G. Eilam and A. Soni,Phys. Rev. D54, 
                  5412  (1996). 
\item[{\rm [16]}] W. Bernreuther, T. Schroder and T. N. Pham, Phys. Lett. B279, 
                389  (1992);\\
                 W. Bernreuther, O. Nachtmann, P. Overmann and T. Schroder,
                                             Nucl. Phys. B388, 53  (1992);\\ 
             W. Bernreuther and A. Brandenburg, Phys. Lett. B314, 104  (1993);\\
	    A. Brandenburg and J. P. Ma, Phys. Lett. B298, 211  (1993);\\
            W. Bernreuther and A. Brandenburg, Phys. Rev. D49, 4481 (1994). 
\item[{\rm [17]}] F. Cuypers and S. D. Rindani, Phys. Lett. B343, 333  (1995). 
\item[{\rm [18]}] D. Atwood and A. Soni,  Phys. Rev. D45, 2405  (1992);\\
                  D. Atwood, G. Eilam, A. Soni, R. Mendel and R. Migneron,
                                      Phys. Rev. Lett. 70, 1364  (1993);\\
            D. Atwood, G. Eilam and A. Soni, Phys. Rev. Lett. 71,492 (1993). 
\item[{\rm [19]}] B. Grzadkowski and J. F. Gunion, 
                     Phys. Lett. B287, 237  (1992);\\
                  R. Cruz, B. Grzadkowski and J. F. Gunion, Phys. Lett. B289, 
                                                                  440 (1992);\\
         N. G. Deshpande, B. Margolis and H. D. Trottier, Phys. Rev. D45, 
                                                               178 (1992);\\
         J. L. Diaz-Cruz and G. Lopez Castro, Phys. Lett. B301, 405 (1993);\\
         C. J. Im, G. L. Kane and P. J. Malde, Phys. Lett. B317, 454  (1993);\\
         B. Grzadkowski and W.-Y. Keung, Phys. Lett. B319, 526 (1993);\\
         D. Chang, W.-Y. Keung and I. Phillips, Phys. Rev. D48,3225  (1993);\\ 
         M. Nowakowski and A. Pilaftsis, Int. J. Mod. Phys. A9, 1097 (1994);\\
         A. Ilakovac, B. A. Kniel and A. Pilaftsis, Phys. Lett. B320, 
                                                                   329 (1994). 
\item[{\rm [20]}] H. L. Lai et. al. , Phys. Rev. D51, 4763 (1995). 
\item[{\rm [21]}] T. Stelzer and S. Willenbrock, Phys. Lett. B357, 125 (1995);\\
                 D. O. Carlson and C.-P. Yuan, hep-ph/9509208, to appear in
                 {\it Proceedings on Physics of the Top Quark}, IITAP, 
                      Iowa State University, Ames, Iowa, May 1995;\\
                 A. P. Heinson, A. S. Belyaev and E. E. Boos, hep-ph/9509274;\\
                 D. Amidei et al. , Fermilab-Pub-96/082. 
\item[{\rm [22]}] C.P.Yuan, Phys.Rev.D41, 42(1990);\\
                  D.Carlson and C.P.Yuan, Phys.Lett.B306,386(1993).

\end{description}
\end{itemize}
\eject

\begin{center}Figure Captions \end{center}

Fig. 1 The asymmetry between the degrees of transverse polarization of the top 
      quark and top antiquark induced by $O_{qW}$ 
       as a function of $\theta_t$ in top pair
      production at the NLC for $\sqrt s=500$ GeV. 

Fig. 2 The asymmetry between the degrees of transverse polarization of the top 
      quark and top antiquark induced by $O_{qW}$ 
      as a function of $\theta_t$ in top pair
      production at the NLC for $\sqrt s=1$ TeV. 

\eject
Table 1 \\
The contribution status of dimension-six CP-violating operators 
to electroweak and $gt\bar t$ couplings. 
The contribution of a CP-violating operator to a particular vertex is
marked by $\times$. 
\vspace{1cm}

\large
\begin{tabular}{|l|c|c|c|c|c|c|c|c|}
\hline
 & & & & & & & & \\ 
  &$~Wt\bar b~$&$~Zt\bar t~$&$~\gamma t\bar t~$&$~Ht\bar t~$
  &$~gt\bar t~$&$~Zb\bar b~$&$~\gamma b\bar b~$&$~Hb\bar b~$ \\ 
 & & & & & & & & \\ \hline

  $~~~O_{t1}$& & & &$\times$& & & & \\ \hline
  $~~~O_{t2}$& &$ $ & &$\times$ & & & & \\ \hline
  $~~~O_{t3}$&$\times $ & & & & & & & \\ \hline
  $~~~O_{Dt}$&$\times $ &$\times $ & 
                      &$\times $ & & & & \\ \hline
  $~~~O_{tW\Phi}$&$\times $&$\times $ &$\times $ & 
      		& & & & \\ \hline
  $~~~O_{tB\Phi}$& &$\times $ &$\times $ & & & & & \\ \hline
  $~~~O_{tG\Phi}$& & & & &$\times$ & & & \\ \hline
  $~~~O_{tG}$& & & & &$\times$ & & & \\ \hline
  $~~~O_{tB}$& &$\times $ &$\times$ & & & & & \\ \hline
  $~~~O_{qG}$& & & & &$\times $& & & \\ \hline
  $~~~O_{qW}$&$\times $ &$\times$ & $\times$ & 
           & &$\times $ &$\times $ & \\ \hline
  $~~~O_{qB}$& &$\times $ &$\times $ & & 
                  &$\times $ &$\times $ & \\ \hline
  $~~~O_{bB}$& & & & & &$\times $ &$\times $ & \\ \hline
  $~~~O_{\Phi q}^{ (1)}$& &$ $ & &$\times$ & & & &$\times$\\ \hline
  $~~~O_{\Phi q}^{ (3)}$&$\times $ & & &$\times$ & 
                    & & &$\times$ \\ \hline
  $~~~O_{\Phi b}$& & & & & & & &$\times$ \\ \hline
  $~~~O_{b1}$& & & & & & & &$\times $ \\ \hline
  $~~~O_{Db}$&$\times$ & & & & &$\times $ & 
           &$\times $ \\ \hline
  $~~~O_{bW\Phi}$&$\times $ & & & & &$\times $
               &$\times$ & \\ \hline
  $~~~O_{bB\Phi}$& & & & & &$\times $ &$\times $ & \\ 
 & & & & & & & & \\ 
\hline
\end{tabular}
\end{document}